\documentclass[aps,prl,twocolumn,showpacs,amsmath,amssymb]{revtex4-2}
\usepackage{epsfig}
\usepackage{graphicx}
\usepackage{dcolumn}
\usepackage{bm}
\usepackage{ltablex,booktabs}
\usepackage{overpic}
\usepackage{subfigure}
\usepackage{float}
\usepackage{color}
\usepackage{amsmath}
\usepackage{mathcomp}
\usepackage{mathrsfs}
\usepackage{multirow}
\usepackage{rotating}
\usepackage{amssymb}
\usepackage{gensymb}
\usepackage{amsmath}
\usepackage{tabularx}
\usepackage[bookmarksnumbered, pdfstartview=FitH,colorlinks,urlcolor=blue, citecolor=blue,linkcolor=blue] {hyperref}

\begin{document}
\normalsize
\parskip=5pt plus 1pt minus 1pt


\title{\boldmath Study of the decay $D_s^+\to K_S^0K_S^0\pi^+$ and observation an isovector partner to $f_0(1710)$ }
\author{
\begin{small}
\begin{center}
M.~Ablikim$^{1}$, M.~N.~Achasov$^{10,b}$, P.~Adlarson$^{68}$, S. ~Ahmed$^{14}$, M.~Albrecht$^{4}$, R.~Aliberti$^{28}$, A.~Amoroso$^{67A,67C}$, M.~R.~An$^{32}$, Q.~An$^{64,50}$, X.~H.~Bai$^{58}$
% ...
\\
\vspace{0.2cm}
(BESIII Collaboration)\\
\vspace{0.2cm} {\it
$^{1}$ Institute of High Energy Physics, Beijing 100049, People's Republic of China\\
% ...
$^{4}$ Bochum Ruhr-University, D-44780 Bochum, Germany\\
% ...
$^{10}$ G.I. Budker Institute of Nuclear Physics SB RAS (BINP), Novosibirsk 630090, Russia\\
% ...
$^{14}$ Helmholtz Institute Mainz, Staudinger Weg 18, D-55099 Mainz, Germany\\
% ...
$^{28}$ Johannes Gutenberg University of Mainz, Johann-Joachim-Becher-Weg 45, D-55099 Mainz, Germany\\
% ...
$^{32}$ Liaoning Normal University, Dalian 116029, People's Republic of China\\
% ...
$^{50}$ State Key Laboratory of Particle Detection and Electronics, Beijing 100049, Hefei 230026, People's Republic of China\\
% ...
$^{58}$ University of Jinan, Jinan 250022, People's Republic of China\\
% ...
$^{64}$ University of Science and Technology of China, Hefei 230026, People's Republic of China\\
% ...
$^{67}$ University of Turin and INFN, (A)University of Turin, I-10125, Turin, Italy; (B)University of Eastern Piedmont, I-15121, Alessandria, Italy; (C)INFN, I-10125, Turin, Italy\\
$^{68}$ Uppsala University, Box 516, SE-75120 Uppsala, Sweden\\
% ...
$^{b}$ Also at the Novosibirsk State University, Novosibirsk, 630090, Russia\\
% ...
}
\end{center}
\vspace{0.4cm}
\end{small}
}
\noaffiliation{}

\date{\today}
\begin{abstract}
  Using $e^+e^-$ annihilation data corresponding to a total integrated
  luminosity of 6.32 $\rm fb^{-1}$ collected at center-of-mass
  energies between 4.178 and 4.226~GeV with the BESIII detector, we
  perform an amplitude analysis of the decay $D_{s}^{+} \to
  K_{S}^{0}K_{S}^{0}\pi^{+}$ for the first time.  An enhancement is
  observed in the $K_{S}^{0}K_{S}^{0}$ mass spectrum near
  1.7~GeV/$c^2$, which was not seen in $D_{s}^{+} \to K^+K^-\pi^{+}$
  in an earlier work, implying the existence of an isospin one partner
  of the $f_0(1710)$.  The branching fraction of the decay $D_{s}^{+}
  \to K_{S}^{0}K_{S}^{0}\pi^{+}$ is determined to be
  $\mathcal{B}(D_{s}^{+} \to K_{S}^{0}K_{S}^{0}\pi^{+})=(0.68\pm0.04_{\rm stat}\pm0.01_{\rm syst})\%$.
\end{abstract}
\maketitle

The constituent quark model has been successful in explaining the
composition of hadrons in the past few decades. In this model, many of
the observed light mesons can be described as $q\bar{q}$ states grouped into
SU(3) flavor multiplets.  In a recent work~\cite{Klempt,
  plb-816-136227}, the $f_0(500)$ and $f_0(980)$ mesons are considered
to be the ground state SU(3) singlet and octet scalar isoscalar
mesons, and the $a_0(980)$ meson is their isovector partner.  The
SU(3) singlet $f_0(1370)$ and octet $f_0(1500)$ are then considered to
be the radial excited states of the $f_0(500)$ and $f_0(980)$ mesons, respectively,
with an isovector partner in the $a_0(1450)$.  However, in case of the
next radial excitation in~\cite{Klempt, plb-816-136227}, for the singlet
$f_0(1710)$ and the newly-identified octet state $f_0(1770)$, no
corresponding isovector $a_0(1710)$ meson has been
established yet.  The BaBar collaboration recently claimed the
observation of a new $a_0(1710)^{\pm}$ resonance in the decay to
$\pi^{\pm}\eta$ with a mass of approximately 1.7~GeV/$c^2$ in the
$\eta_c \to \eta \pi^+\pi^-$ decay~\cite{a01710}.  Constructing
isospin eigenstates from kaon pairs, we obtain ($\left|K^+K^-\right>-\left|K^0\bar{K}^0\right>$)
for isospin one, but ($\left|K^+K^-\right> + \left|K^0\bar{K}^0\right>$) for isospin zero.  It
follows that if the interference between an $f_0$ and an $a_0$ is
constructive in decays to a $K^+K^-$ pair, it is destructive in decays
to a pair of neutral kaons and vice versa.  A comparison between
decays involving $K^+K^-$ and $K_S^0 K_S^0$ pairs can thus give access
to such interference terms and allows a search for the $a_0(1710)^0$
in decays to two kaons.

The BESIII and the BaBar collaborations reported analyses of the
$D_s^+\to K^+K^-\pi^+$ decay~\cite{MWang, prd-83-052001} and observed
contributions of the scalar mesons $S(980)$ (where $S(980)$ denotes an
admixture of $a_0(980)^0$ and $f_0(980)$) and $f_0(1710)$.  Both
collaborations reported consistent results for the branching fractions~(BF)
$\mathcal{B}(D_s^+\to S(980)\pi^+, S(980)\to K^+K^-) = (1.05\pm0.04_{\rm stat}\pm 0.06_{\rm syst})\%$ and
$\mathcal{B}(D_s^+\to f_0(1710)\pi^+, f_0(1710)\to K^+K^-) = (0.10\pm 0.02_{\rm stat}\pm 0.03_{\rm syst})\%$~\cite{MWang}.

Furthermore, analyses of $D_s$ decays are an important input for studies of the
$B^0_s$ meson, which predominantly decays to $D_s+X$~\cite{PDG}.  In addition,
hadronic $D_s$ decays probe the interplay of short-distance weak-decay matrix
elements and long-distance QCD interactions. The measurement of BFs of hadronic
$D_s$ decays provides valuable information to help understand strong force-induced
amplitudes and phases~\cite{PRD79-034016, PRD81-074021, PRD84-074019, BCKa0}.

The CLEO collaboration measured the absolute BF of the decay
$D_{s}^{+} \to K_{S}^{0}K_{S}^{0}\pi^{+}$ to be
$(0.77\pm 0.05_{\rm stat}\pm0.03_{\rm syst})\%$~\cite{CLEO-BF}, using a dataset
corresponding to a luminosity of 586 pb$^{-1}$ at a center-of-mass energy of
4.17~GeV.  In this work, we present the first amplitude analysis and a more
precise measurement of the BF of the $D_{s}^{+} \to K_{S}^{0}K_{S}^{0}\pi^{+}$
decay using 6.32~$\rm fb^{-1}$ of data samples collected at center-of-mass
energies of 4.178, 4.189, 4.199, 4.209, 4.219 and 4.226~GeV with the BESIII
detector.  We do not distinguish between the $a_0(1710)^{0}$ and $f_0(1710)$
mesons, and denote the combined state as $S(1710)$.  Charge conjugation is
implied throughout this paper.

The BESIII detector~\cite{Ablikim:2009aa, Ablikim:2019hff} records
symmetric $e^+e^-$ collisions provided by the BEPCII storage
ring~\cite{Yu:IPAC2016-TUYA01}.  The cylindrical core of the BESIII
detector covers 93\% of the full solid angle and consists of a
helium-based multilayer drift chamber~(MDC), a plastic scintillator
time-of-flight system~(TOF), and a CsI(Tl) electromagnetic
calorimeter~(EMC), which are all enclosed in a superconducting
solenoidal magnet providing a 1.0~T magnetic field.  The end cap TOF
system was upgraded in 2015 using multi-gap resistive plate chamber
technology~\cite{etof}.

Simulated data samples produced with {\sc geant4}-based~\cite{geant4} Monte
Carlo (MC) software, which includes the geometric description of the BESIII
detector and the detector response, are used to determine detection
efficiencies and to estimate backgrounds. The simulation models the beam energy
spread and initial state radiation (ISR) in the $e^+e^-$ annihilations with the
generator {\sc kkmc}~\cite{ref:kkmc}. The inclusive MC sample includes the
production of open charm processes, the ISR production of vector
charmonium(-like) states, and the continuum processes incorporated in
{\sc kkmc}~\cite{ref:kkmc}. The known decay modes are modeled with
{\sc evtgen}~\cite{ref:evtgen} using BFs taken from the Particle Data
Group~\cite{PDG}, and the remaining unknown charmonium decays are modeled with
{\sc lundcharm}~\cite{ref:lundcharm}. Final state radiation~(FSR) from charged
final state particles is incorporated using {\sc photos}~\cite{photos}.

The process
$e^{+}e^{-} \to D_{s}^{*\pm}D_{s}^{\mp}\to \gamma D_{s}^{+}D_{s}^{-}$
allows studies of $D_s^{+}$ decays with a tag technique~\cite{MarkIII-tag}.
There are two types of samples used in the tag technique: single tag~(ST) and
double tag~(DT). In the ST sample, a $D_{s}^{-}$ meson is reconstructed through
a particular hadronic decay without any requirement on the remaining measured
tracks and EMC showers. In the DT sample, a $D_{s}^{+}$, designated as the
``signal'', is reconstructed through $D_{s}^{+} \to K^0_{S}K^0_{S}\pi^{+}$,
while a $D_{s}^{-}$, designated as ``tag'', is reconstructed through one of
eight hadronic decay modes: $D_s^-\to K_{S}^{0}K^{-}$, $K^{+}K^{-}\pi^{-}$,
$K^{+}K^{-}\pi^{-}\pi^{0}$, $K_{S}^{0}K^{-}\pi^{-}\pi^{+}$,
$K_{S}^{0}K^{+}\pi^{-}\pi^{-}$, $\pi^{-}\pi^{-}\pi^{+}$, $\pi^{-}\eta^{\prime}$,
and $K^{-}\pi^{-}\pi^{+}$. A detailed description of selection conditions
concerning charged and neutral particle candidates, the mass recoiling against
$D_s^{\pm}$ candidates, and the mass of the tag candidates are provided in
Refs.~\cite{ref:a0980, ref:Kspipi0, ref:KsKpipi}.

As in Refs.~\cite{ref:Kspipi0, ref:KsKpipi}, an eight-constraint (8C)
kinematic fit is performed to select signal events for the amplitude
analysis. Besides the
constraints arising from four-momentum conservation, the invariant
masses of the two $K_S^0$ candidates, the tag $D_s^-$, and the
$D_s^{*+(-)}$ candidates are constrained to their known masses given
in Ref.~\cite{PDG}. If there are multiple signal combinations, the
candidate with the minimum $\chi^2$ of the 8C kinematic fit is
chosen. Signal $D_s^+$ candidates are selected if their invariant mass is in
the interval [1.950, 1.990]~GeV/$c^2$. A further kinematic fit including
a ninth constraint on the mass of the signal $D_s^+$ is performed, and
the updated four-momenta are used for the amplitude analysis. This
ensures that all candidates fall within the phase space boundary. In
total, 412 events are selected with a purity of $f_s =
(97.3\pm0.8)$\%.  The purity is determined from a fit to the invariant
mass distribution of the signal $D_s^+$ candidates.

This analysis uses an isobar formulation in the covariant tensor formalism~\cite{covariant-tensors}.
The total amplitude $M$ for the decay is described by a coherent sum of the amplitudes
of all intermediate processes, $M=\begin{matrix}\sum_n c_{n}A_{n}\end{matrix}$, where
$n$ indicates the $n^{\rm th}$
intermediate state and $c_{n}=\rho_{n}e^{i\phi_{n}}$ is the corresponding
complex coefficient with magnitude $\rho_{n}$ and phase $\phi_{n}$. The model is symmetrized
with respect to the two
identical $K_S^0$ mesons. The two-body decay amplitude $A_{n}$ is
given by $A_{n} = P_{n}S_{n}F_{n}^{r}F_{n}^{D}$, where $S_{n}$ and
$F_{n}^{r(D)}$ are the spin factor~\cite{covariant-tensors} and the
Blatt-Weisskopf barrier factor of the intermediate state (the $D_{s}^{\pm}$
meson)~\cite{Blatt}, respectively, and $P_{n}$ is the relativistic Breit-Wigner
amplitude~\cite{RBW} describing the propagator of the intermediate
resonance.

Contributions of intermediate resonances are determined by an unbinned
maximum-likelihood fit to data.
A combined probability density
function~(PDF) for the signal and background hypotheses is constructed, depending on the momenta
of the three final-state particles. See Refs.~\cite{ref:Kspipi0, ref:KsKpipi, prd-85-122002} for details.
The signal PDF is constructed from the total
amplitude $M$. The background PDF, $B$, is constructed from a background shape
derived from the inclusive MC samples using the kernel estimation method RooNDKeysPdf~\cite{Verkerke,Cranmer}.
It models the distribution of an input dataset as a
superposition of Gaussian kernels. This background PDF is then added to the
signal PDF incoherently and the combined PDF is written as
\begin{eqnarray}
  \begin{aligned}
    \textrm{PDF} = \epsilon R_3&\left[\frac{f_s\left|M(p_{j})\right|^{2}}{\int \epsilon\left|M(p_{j})\right|^{2}R_3\,dp_{j}}
      +\frac{(1-f_s)B_{\epsilon}(p_{j})}{\int \epsilon B_{\epsilon}(p_{j})R_3\,dp_{j}}\right], \nonumber 
  \end{aligned}
\end{eqnarray}
where $B_{\epsilon}$ is defined as $B/\epsilon$, $\epsilon$ is the acceptance in bins of the Dalitz plot determined
with a MC sample of $D_s^+\to K_S^0K_S^0\pi^+$ uniformly distributed over the Dalitz plot. The
placeholder $p_j=\{p_1,\;p_2,\;p_3\}$ represents the momenta of the final state particles, $R_3$ is the three-body phase-space element,
and $f_s$ is the purity.
The normalization integral
in the denominator is determined by a MC technique as described in
Ref.~\cite{ref:Kspipi0, ref:KsKpipi}.

\begin{figure}[!htp]
  \centering
  \includegraphics[width=0.235\textwidth]{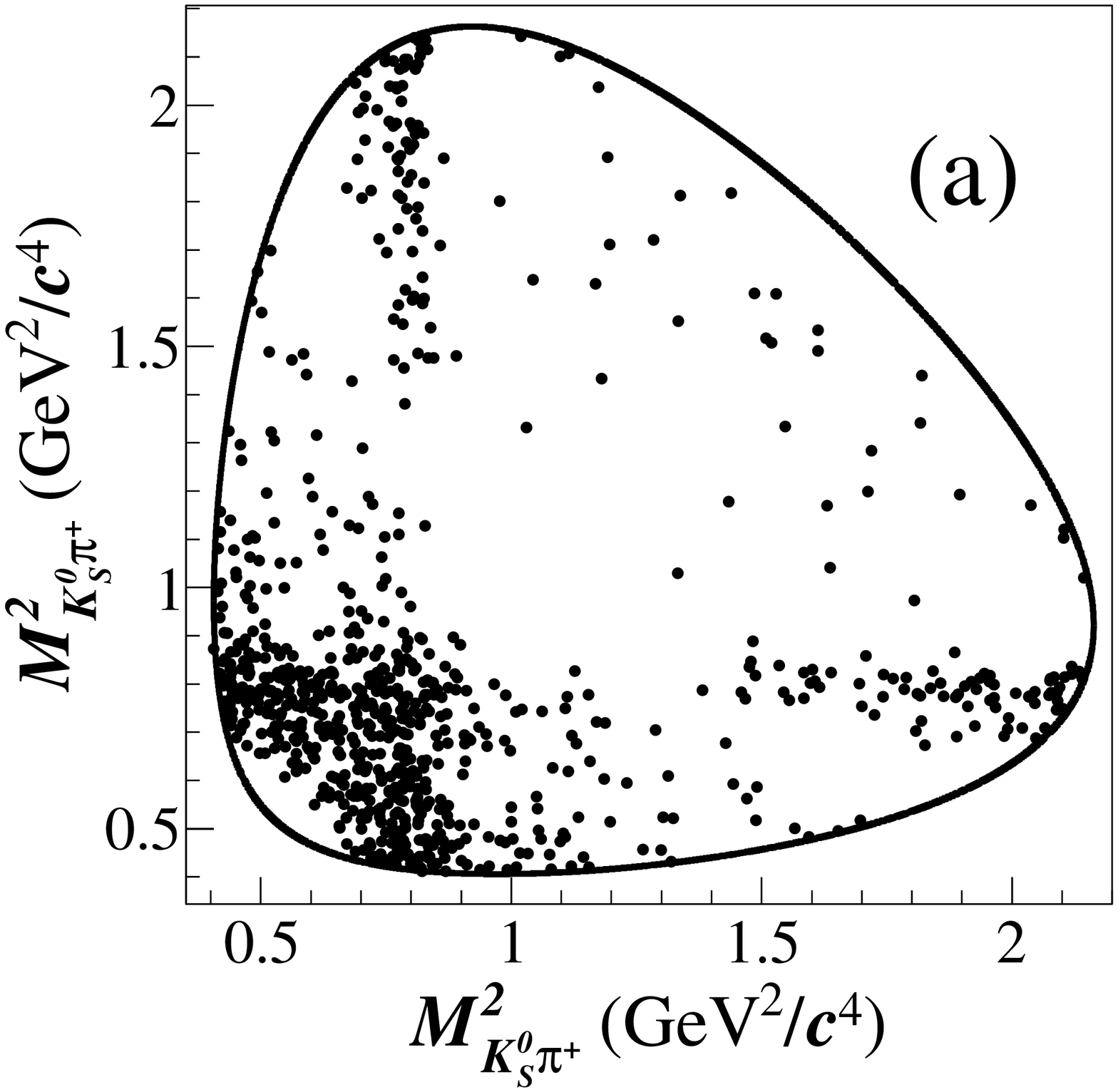}
  \includegraphics[width=0.235\textwidth]{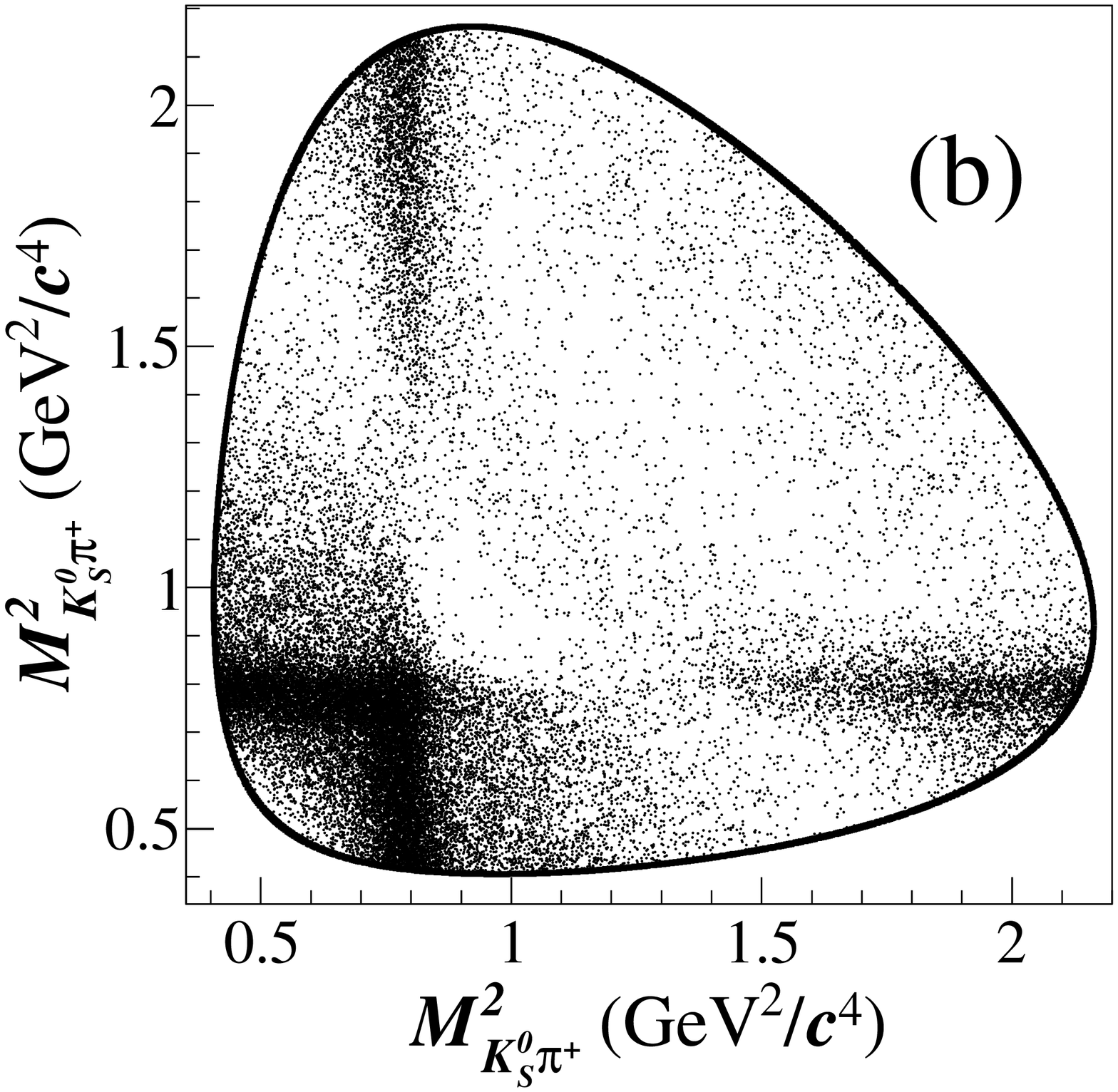}
    \caption{Dalitz plot of $M^{2}_{K_S^0\pi^+}$ versus~$M^{2}_{K_S^0\pi^+}$
      for $D^+_s\to K^0_SK^0_S\pi^+$, symmetrized for the indistinguishable
      $K_S^0$ candidates (two entries per event), of (a) the sum of all data samples and (b) the signal MC samples generated
      based on the amplitude analysis result. The black curve indicates the
      kinematic boundary.}
    \label{fig:dalitz}
\end{figure}

\begin{figure}[!htbp]
  \centering
  \includegraphics[width=0.235\textwidth]{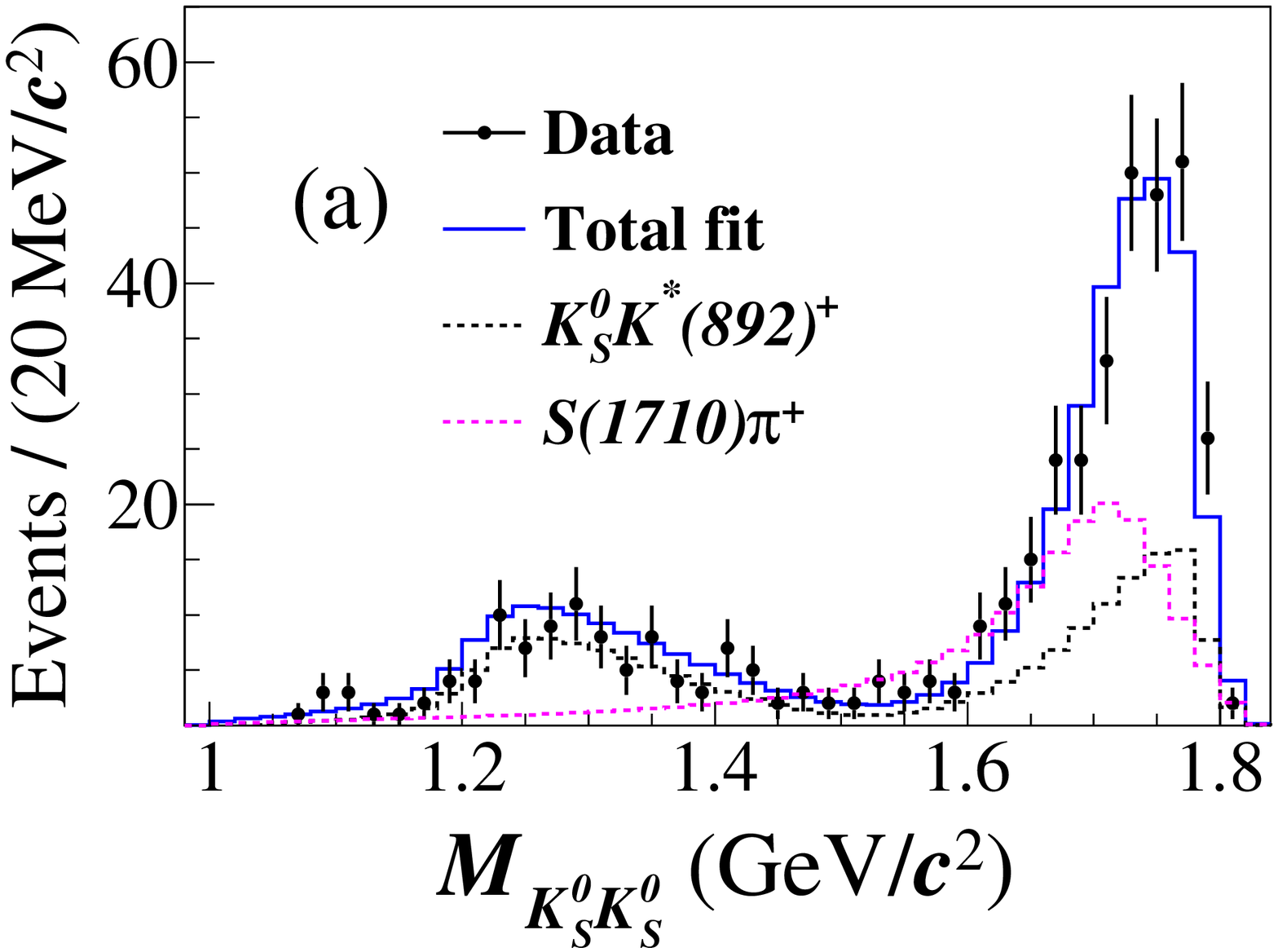}
  \includegraphics[width=0.235\textwidth]{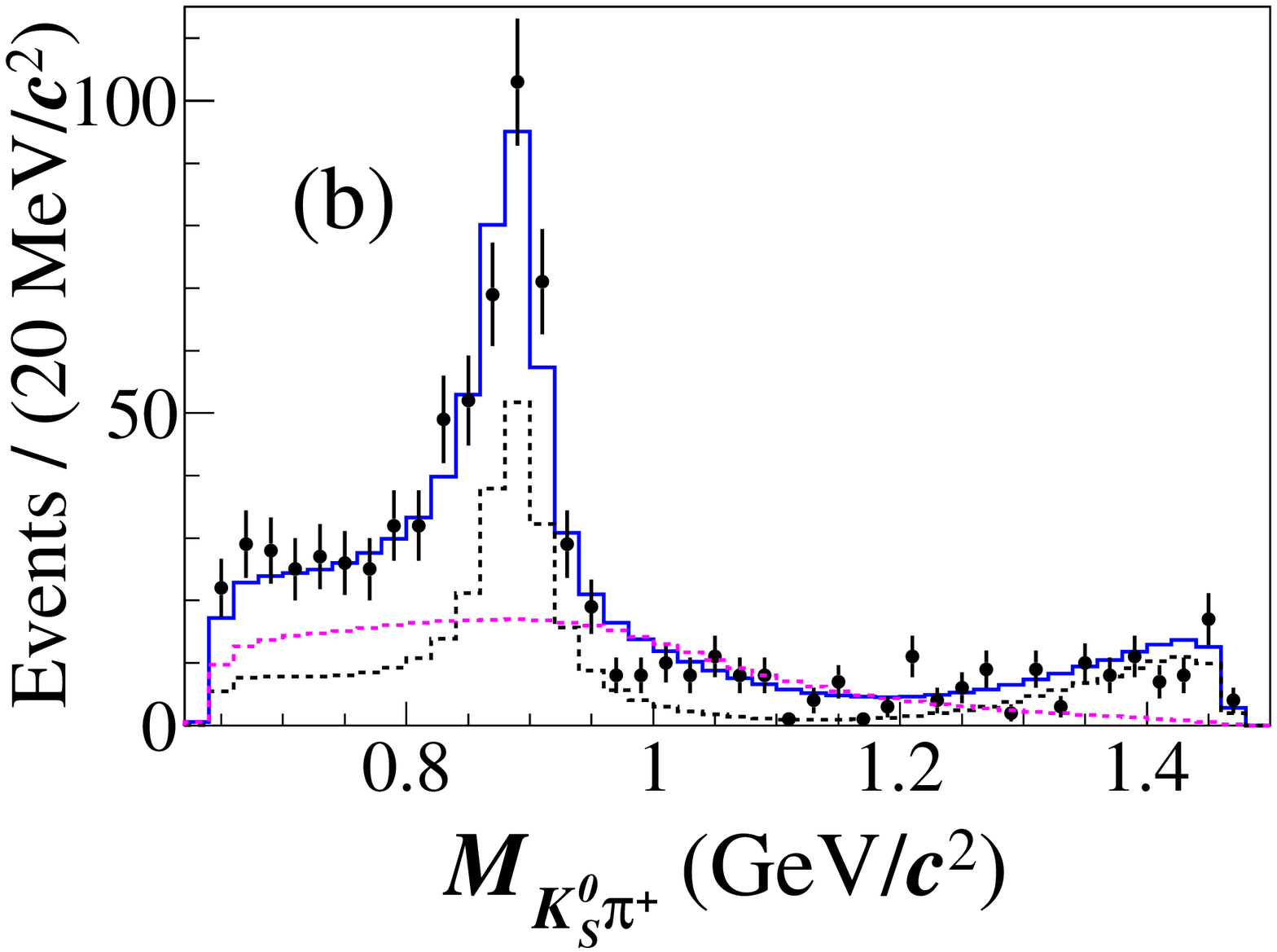}
  \caption{
    Distribution of (a) $M_{K_S^0K_S^0}$ and (b) $M_{K_S^0\pi^+}$ from the
    nominal fit. The distribution of $M_{K_S^0\pi^+}$ contains two entries per event, one for each $K_S^0$.
    The data samples
    are represented by points with uncertainties and the fit results by the blue
    lines. Colored dashed lines show the individual components of the fit model.
    Due to interference effects, the total PDF is not necessarily equal to the sum
    of the components.}
  \label{dalitz-projection}
\end{figure}

The Dalitz plot of $M^2_{K_S^0\pi^+}$ versus $M^2_{K_S^0\pi^+}$ is shown in
Fig.~\ref{fig:dalitz}(a). The strong vertical and horizontal bands around
0.8~GeV$^2$/$c^4$ are caused by the process $D_s^+\to K_S^0K^{*}(892)^+$.
We choose this process as a reference so that the magnitudes and phases of other amplitudes are to be understood
as relative values with respect to this reference amplitude.
The purity is a fixed quantity in the fit.
Other possible contributions from resonances such as $K_{1}(1410)^{+}$,
$K^{*}_{0}(1430)^{+}$, $a_{0}(980)$, $f_{0}(980)$, $f_{2}(1270)$, $a_{2}(1320)$,
$f_{0}(1370)$, $a_{0}(1450)$, $f_{0}(1500)$, $f_{2}(1525)$, $a_{2}(1700)$ and $S(1710)$ are added to the
fit one at a time.
The masses and widths of all resonances
are fixed to the known values~\cite{PDG} apart from those of the $S(1710)$.
The statistical significance of each new amplitude
is calculated from the change of the log-likelihood taking the change in the number of
degrees of freedom into account. Various combinations of these resonances
are also tested. In addition to the reference amplitude $D_s^+\to K_S^0K^{*}(892)^+$,
the amplitude for the decay $D_s^+\to S(1710)\pi^+$ is found to have a significance larger than $10\sigma$.
No other contribution has a significance of more than $3\sigma$.
The significance of a $a_{0}(980)/f_{0}(980)$ contribution is less than $0.1\sigma$.
The Dalitz plot of the signal MC sample
generated based on the result of the amplitude analysis is shown in Fig.~\ref{fig:dalitz}(b).
The mass projections of the fit are shown in Fig.~\ref{dalitz-projection}. The
goodness of fit is $\chi^{2}/\rm {NDOF} = 15.9/19 = 0.8$ for
Fig.~\ref{dalitz-projection}(a) and $28.8/32 = 0.9$ for
Fig.~\ref{dalitz-projection}(b), where $\rm NDOF$ is the number of degrees of
freedom. In the goodness of fit calculation, we merge neighboring bins until each bin
has at least 10 entries.

The contribution of the $n$th amplitude relative to the total BF is quantified
by the fit fraction~(FF) defined as
${\rm FF}_{n} = \int \left|\rho_{n}A_{n}\right|^{2}dR_3/\int\left|M\right|^{2}dR_3$.
The FFs for both amplitudes and the phase difference relative to the reference process are listed in Table~\ref{fit-result}.
The sum of the two FFs is 89.8$\%$.
The Breit-Wigner mass and width of the $S(1710)$ are determined to be
$(1.723\pm0.011_{\rm stat}\pm0.002_{\rm syst})$~GeV/$c^{2}$ and
$(0.140 \pm 0.014_{\rm stat} \pm 0.004_{\rm syst})$~GeV/$c^{2}$, respectively.
\begin{table}[!htbp]
  \caption{Fit Fractions (FF) for the two amplitudes, and phase difference to the reference process. The first and the second
    uncertainties are statistical and systematic, respectively. The sum of the two FFs is
    89.8$\%$.}
  \label{fit-result}
  \begin{center}
    \begin{tabular}{lccc}
      \hline \hline
      Amplitude                                 & Phase                            & FF~(\%)                \\
      \hline
      $D_{s}^{+} \to K_{S}^{0}K^{*}(892)^{+}$   & 0.0(fixed)                       & $43.5 \pm 3.9 \pm 0.5$  \\
      $D_{s}^{+} \to S(1710)\pi^{+}$        & \phantom{0}$2.3 \pm 0.1 \pm 0.1$ & $46.3 \pm 4.0 \pm 1.2$  \\
      \hline \hline
    \end{tabular}
  \end{center}
\end{table}

Systematic uncertainties for the results of the amplitude analysis, including the
phase difference, FFs, and the mass and the width of the $S(1710)$, are determined by
differences between the results of the nominal fit and fits with
the following variations. The mass and the width of the $K^{*}(892)^{+}$ are
shifted by their uncertainties~\cite{PDG}. The radii of the Blatt-Weisskopf
barrier factors are varied from their nominal values of $5$~GeV$^{-1}$ and $3$~GeV$^{-1}$ (for the $D_s^+$ meson
and the intermediate resonances, respectively)  by $\pm 1$~GeV$^{-1}$. The uncertainties associated with the
size of the background sample are studied by varying the purity within its statistical
uncertainty. An alternative background sample is used to determine the background PDF, where
the relative fractions of background processes from direct $q\bar{q}$ and
non-$D_{s}^{*\pm}D_{s}^{\mp}$ open-charm processes are varied by the statistical uncertainties of the known cross sections.
To estimate the systematic
uncertainty related to the reconstruction efficiency, the amplitude analysis is
performed varying the particle-identification and tracking efficiencies
according to their uncertainties. The total uncertainties are obtained by
adding these contributions in quadrature.

\begin{figure}[htp]
  \begin{center}
    \includegraphics[width=0.48\textwidth]{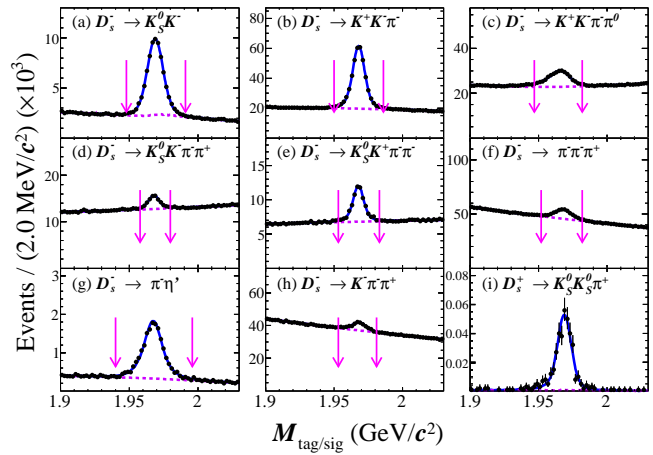}
    \caption{Fits to (a)-(h) the $M_{\rm tag}$ distributions of the ST
      candidates and (i) the $M_{\rm sig}$ distribution of the DT
      signal candidates. The data samples are represented by points with uncertainties,
      the total fit results by solid blue lines and the background
      contributions by dashed violet lines. The pairs of pink arrows indicate the signal regions.}
    \label{fig:DT_fit}
  \end{center}
\end{figure}

The BF of $D_s^+\to K_S^0K_S^0\pi^+$ is measured with the DT technique
using the same tag modes and event selection criteria as in the
amplitude analysis.  However, the kinematic fit is not applied.  We
require the momentum of the isolated $\pi^+$ to be greater than
$>0.1$~GeV/$c$ to remove soft pions from $D^{*+}$ decays.  For each
tag mode, the best ST candidate of the tag $D_s^-$ is chosen as the
combination with the recoiling mass closest to the known $D_s^{*+}$
mass~\cite{PDG} and the best DT candidate is chosen as the combination
with the average mass of the tag $D_s^-$~($M_{\rm tag}$) and the
signal $D_s^+$~($M_{\rm sig}$) closest to the $D_s^{*+}$ mass. The BF
is given by~\cite{ref:Kspipi0, ref:KsKpipi}
\begin{eqnarray} \begin{aligned}
    \mathcal{B}_{\text{sig}}=\frac{N_{\text{total,sig}}^{\text{DT}}}{\sum_{\alpha,   
        i} N_{\alpha, i}^{\text{ST}}\epsilon^{\text{DT}}_{\alpha,\text{sig},
        i}/\epsilon_{\alpha, i}^{\text{ST}}},\, \label{eq:Bsig-gen}
\end{aligned} \end{eqnarray} 
where $\alpha$ runs over the various tag modes, and $i$ denotes the
different center-of-mass energies.  The ST yields in data $N_{\alpha,
  i}^{\text{ST}}$ and the DT yield $N_{\text{total,sig}}^{\text{DT}}$
are determined by fits to the $M_{\rm tag}$ and $M_{\rm sig}$
distributions shown in Figs.~\ref{fig:DT_fit}(a-f) and
Fig.~\ref{fig:DT_fit}(i), respectively. The signal shape is modeled
with the MC-simulated shape convolved with a Gaussian function. In the
fits to the $M_{\rm tag}$ distributions, the background is
parameterized as a second-order Chebyshev polynomial. For the tag
modes $D_{s}^{-} \to K_{S}^{0}K^{-}$ and $D_{s}^{-} \to
\pi^{-}\eta^{\prime}$, MC simulations of the decays $D^{-} \to
K_{S}^{0} \pi^-$ and $D_{s}^{-} \to \eta\pi^+\pi^-\pi^-$ are added to
the background to account for these peaking background contributions.
In the fit to the $M_{\rm sig}$ distribution, the background is
described by the background MC. The corresponding efficiencies
$\epsilon$ are obtained by analyzing the inclusive MC samples, with
the signal events for $D_s^+\to K_S^0K_S^0\pi^+$ generated based on
the results of the amplitude analysis. The total ST yields of all tag
modes and the DT yields are $531217\pm2235$ and $371\pm21$,
respectively. The BF of $D_s^+\to K_S^0K_S^0\pi^+$ is determined to be
$(0.68\pm0.04_{\rm stat}\pm0.01_{\rm syst})\%$.

We consider the following systematic uncertainties in the measurement of the BF.
Varying the signal and background shapes and taking into account the background fluctuation,
the uncertainty on the total number of ST $D_s^-$ candidates is 0.4\%.

The uncertainty associated with the background shape in the fit to the
DT $M_{\rm sig}$ distribution is 0.3\%, determined by replacing the
nominal background shape with a second-order Chebyshev function and
taking the difference between the two results.  The uncertainty of the
$K_{S}^{0}$ reconstruction efficiency is examined using control
samples of $J/\psi\to K_{S}^{0}K^{\pm}\pi^{\mp}$ and $\phi
K_{S}^{0}K^{\pm}\pi^{\mp}$ decays, and the data-MC efficiency ratio is
$(101.01\pm0.53)\%$~\cite{prd-92-112008}.  We correct the signal
efficiencies by this factor, and use the uncertainty of 0.53\% as a
systematic uncertainty.  The $\pi^{+}$ tracking efficiencies are
studied with $e^+e^-\to K^+K^-\pi^+\pi^-$ events. The data-MC
efficiency differences of the $\pi^{+}$ particle-identification and
tracking are both 1.0\%.  The uncertainty from the signal MC based on
the results of the amplitude analysis is studied by varying the fit
parameters according to the covariance matrix. The change of signal
efficiency is estimated to be 0.5\%. The uncertainty due to the
limited MC sample size is obtained from
$\sqrt{\sum_{\alpha,i}(f_{\alpha,
    i}\frac{\delta_{\epsilon_{\alpha,i}}}{\epsilon_{\alpha,i}}})^2$,
where $f_{\alpha,i}$ is the tag yield fraction, and $\epsilon_{i}$ and
$\delta_{\epsilon_{i}}$ are the signal efficiency and the
corresponding uncertainty of tag mode $\alpha$ and center-of-mass
energy $i$, respectively. It is found to be 0.2\%.  In total, the
systematic uncertainty on the branching fraction is 1.9\%.

In summary, we present the first amplitude analysis of the decay
$D_{s}^{+} \to K_{S}^0K_{S}^0\pi^{+}$ using 6.32 fb$^{-1}$ of $e^+e^-$
annihilation data taken at center-of-mass energies between 4.178 and
4.226~GeV. The results are listed in Table~\ref{fit-result}.
The Breit-Wigner mass and width of the $S(1710)$ are measured
to be  ($1.723 \pm 0.011_{\rm stat} \pm 0.002_{\rm syst}$)~GeV/$c^2$ and
($0.140 \pm 0.014_{\rm stat} \pm 0.004_{\rm syst}$)~GeV/$c^2$, respectively.
These parameters are consistent with the PDG evaluation for the $f_0(1710)$
within 1.2$\sigma$ and 0.7$\sigma$, respectively~\cite{PDG}.

 
The BF of $D_{s}^{+} \to K_{S}^0K_{S}^0\pi^{+}$ is determined to be
$(0.68\pm0.04_{\rm stat}\pm 0.01_{\rm syst})\%$, which is consistent with
the CLEO result
$\mathcal{B}(D_{s}^{+} \to K_{S}^0K_{S}^0\pi^{+})= (0.77\pm 0.05_{\rm stat} \pm 0.03_{\rm syst})\%$~\cite{PDG, CLEO-BF}
within $1.3\sigma$. The BFs for the two intermediate processes are calculated with
$\mathcal{B}_{i} = {\rm FF}_{i} \times \mathcal{B}(D_{s}^{+} \to K_{S}^0K_{S}^0\pi^{+})$,
as shown in Table~\ref{inter-processes}. The BF of
$D_{s}^{+} \to K_{S}^0K^{*}(892)^{+}, K^{*}(892)^{+} \to K_{S}^0\pi^+$ is
determined to be $(3.0\pm 0.3_{\rm stat} \pm 0.1_{\rm syst})\times10^{-3}$.
This leads to
$\mathcal{B}(D_{s}^{+} \to \bar{K^0}K^{*}(892)^{+})= (1.8 \pm 0.2_{\rm stat} \pm 0.1_{\rm syst})\%$
which deviates from the CLEO result of this BF, $(5.4\pm 1.2)\%$ ~\cite{PDG, plb-226-192}, by $2.9\sigma$.
However, Ref.~\cite{plb-226-192} does not consider interference terms.

Because a significant $D_{s}^{+} \to f_0(980)/a_0(980)^0\pi^+$
contribution is observed in the amplitude analysis of $D^+_s\to
K^+K^-\pi^+$~\cite{MWang}, one would expect that about $10\%$ of the
signal comes from $D_{s}^{+} \to f_0(980)/a_0(980)^0\pi^+$ with
$f_0(980)/a_0(980)^0 \to K_{S}^0K_{S}^0$~\cite{PDG}.  However, almost
no signal populates the region below 1.1~GeV/$c^2$ in the
$K_{S}^0K_{S}^0$ mass spectrum.  This suppression can likely be
attributed to destructive interference between $a_0(980)^0$ and
$f_0(980)$ in decays to two neutral kaons. The same interference term
would then be constructive in decays to two charged kaons, explaining
the large branching fraction observed there.  On the other hand, an
enhancement is seen in the $K_S^{0}K_S^{0}$ mass spectrum around
1.7~GeV/$c^2$.  Ref.~\cite{MWang} reports $\mathcal{B}(D_{s}^{+} \to
f_0(1710)\pi^{+}, f_0(1710)\to K^{+}K^{-})= (0.10\pm 0.02_{\rm stat}
\pm 0.03_{\rm syst})\%$.  This corresponds to an expected BF of about $5\times
10^{-4}$ for $D_{s}^{+} \to f_0(1710)\pi^{+}, f_0(1710)\to
K_S^{0}K_S^{0}$, based on isospin symmetry predicting the ratio
$\frac{\mathcal{B}(f_{0}(1710)\to
  K^{+}K^{-})}{\mathcal{B}(f_0(1710)\to K_{S}^0K_{S}^0)}$ to be two.
In our amplitude analysis, we determine this BF to be $(3.1 \pm
0.3_{\rm stat} \pm 0.1_{\rm syst})\times 10^{-3}$, which is one order
of magnitude larger than the expectation.  Based on the same argument
concerning the difference in interference between pairs of charged and
neutral kaons in isospin one and isospin zero configurations, this
observation implies the existence of an isospin one partner of the
$f_0(1710)$ meson, the $a_0(1710)^{0}$, as proposed by Ref.~\cite{Klempt}
and as recently observed in Ref.~\cite{a01710}.  The $f_0(1710)$ and
$a_0(1710)^{0}$ amplitudes could then interfere constructively in
decays to two neutral kaons and destructively in decays to two charged
kaons, explaining the different observations made in this work and in
Ref.~\cite{MWang}.  A simultaneous amplitude analysis of $D_s^+\to
K^+K^-\pi^+$ and $D_s^+\to K_S^0K_S^0\pi^+$ can further clarify this
situation. In addition, a charged partner of $a_0(1710)^{0}$ is
expected to be visible in the $K_S^0K^+$ mass spectrum in the related decay
$D_s^+\to K_S^0K^+\pi^0$~\cite{DsKsKppiz}. 


\begin{table}[htbp]
  \caption{BFs for amplitudes with the final state
    $K_{S}^0K_{S}^0\pi^{+}$. The first and the second uncertainties are
    statistical and systematic, respectively.}\label{inter-processes}
  \begin{center}
    \begin{tabular}{lcc}
      \hline\hline
      Amplitude                     & BF ($10^{-3}$)  \\
      \hline
      $D_{s}^{+} \to K_{S}^{0}K^{*}(892)^{+}\to K_{S}^{0}K_{S}^{0}\pi^+$  & $ 3.0 \pm 0.3 \pm 0.1$  \\
      $D_{s}^{+} \to S(1710)\pi^{+}\to K_{S}^{0}K_{S}^{0}\pi^+$           & $ 3.1 \pm 0.3 \pm 0.1$  \\
      \hline\hline
\end{tabular}
\end{center}
\end{table}

\acknowledgements
The BESIII collaboration thanks the staff of BEPCII and the IHEP computing center for their strong support. This work is supported in part by National Key R\&D Program of China under Contracts Nos. 2020YFA0406400, 2020YFA0406300; National Natural Science Foundation of China (NSFC) under Contracts Nos. 11625523, 11635010, 11735014, 11822506, 11835012, 11875054, 11935015, 11935016, 11935018, 11961141012, 12022510, 12025502, 12035009, 12035013, 12061131003; the Chinese Academy of Sciences (CAS) Large-Scale Scientific Facility Program; Joint Large-Scale Scientific Facility Funds of the NSFC and CAS under Contracts Nos. U2032104, U1732263, U1832207; CAS Key Research Program of Frontier Sciences under Contract No. QYZDJ-SSW-SLH040; 100 Talents Program of CAS; INPAC and Shanghai Key Laboratory for Particle Physics and Cosmology; ERC under Contract No. 758462; European Union Horizon 2020 research and innovation programme under Contract No. Marie Sklodowska-Curie grant agreement No 894790; German Research Foundation DFG under Contracts Nos. 443159800, Collaborative Research Center CRC 1044, FOR 2359, GRK 2149; Istituto Nazionale di Fisica Nucleare, Italy; Ministry of Development of Turkey under Contract No. DPT2006K-120470; National Science and Technology fund; Olle Engkvist Foundation under Contract No. 200-0605; STFC (United Kingdom); The Knut and Alice Wallenberg Foundation (Sweden) under Contract No. 2016.0157; The Royal Society, UK under Contracts Nos. DH140054, DH160214; The Swedish Research Council; U. S. Department of Energy under Contracts Nos. DE-FG02-05ER41374, DE-SC-0012069.

\end{document}